\documentclass[a4paper, amsfonts, amssymb, amsmath, reprint, prl, aps,
showkeys, nofootinbcib,superscriptaddress,twocolumn,twoside]{revtex4-2}
\usepackage{lipsum}
\usepackage{amsthm}
\usepackage{mathtools}
\usepackage{physics}
\usepackage{xcolor}\usepackage{soul}
\usepackage{graphicx}
\usepackage{textcomp}
\usepackage{epstopdf}

\usepackage[normalem]{ulem}  
\usepackage{adjustbox}
\usepackage{placeins}
\usepackage[T1]{fontenc}
\usepackage{lipsum}
\usepackage[english]{babel}
\usepackage[utf8]{inputenc}
\usepackage{csquotes}
\usepackage[mathscr]{eucal} 
\usepackage{bibunits} 
\usepackage{color}
\usepackage[unicode=true,
bookmarks=true,bookmarksnumbered=false,bookmarksopen=false,
breaklinks=false,pdfborder={0 0 1},backref=false,colorlinks=true]
{hyperref}
\hypersetup{
	colorlinks=true,
	linkcolor=blue,
	citecolor=blue,
	urlcolor=blue,
	pdfstartview={FitH}
}

\usepackage[capitalize]{cleveref}
\usepackage{multirow}

\newcommand{\orcid}[1]{\href{https://orcid.org/#1}{\includegraphics[width=6.6pt]{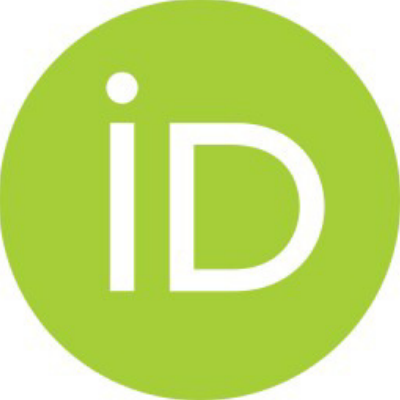}}}

\definecolor{darkblue}{rgb}{0.0, 0.0, 0.55}

\begin{document}
\title{Quantum Error Correction with Superpositions of Squeezed Fock States}
\author{Yexiong Zeng~\orcid{0000-0001-9165-3995}}

\affiliation{RIKEN Center for Quantum Computing, RIKEN, Wakoshi, Saitama 351-0198, Japan}
\affiliation{Key Laboratory of Low-Dimensional Quantum Structures and Quantum Control of Ministry of Education,
	Department of Physics and Synergetic Innovation Center for Quantum Effects and Applications,
	Hunan Normal University, Changsha 410081, China}

\author{Fernando Quijandría~\orcid{0000-0002-2355-0449}}\affiliation{RIKEN Center for Quantum Computing, RIKEN, Wakoshi, Saitama 351-0198, Japan}	

\author{Clemens Gneiting~\orcid{0000-0001-9686-9277}}
\altaffiliation[clemens.gneiting@riken.jp]{}
\affiliation{RIKEN Center for Quantum Computing, RIKEN, Wakoshi, Saitama 351-0198, Japan}

\author{Franco Nori~\orcid{0000-0003-3682-7432}}
\altaffiliation[fnori@riken.jp]{}
\affiliation{RIKEN Center for Quantum Computing, RIKEN, Wakoshi, Saitama 351-0198, Japan}
\affiliation{Quantum Research Institute and Department of Physics, University of Michigan, Ann Arbor, Michigan, 48109-1040, USA}

\date{\today}

\begin{abstract}
Bosonic codes, leveraging infinite-dimensional Hilbert spaces for redundancy, offer great potential for {encoding quantum information}. {However, a practical continuous-variable bosonic code that can \textit{simultaneously correct both photon loss and dephasing errors}, while achieving a high level of compliance with the Knill--Laflamme conditions within an experimentally friendly structure, remains elusive.} Here, we propose a {code based on the superposition of squeezed Fock states} with an error-correcting capability that scales as $\propto\exp(-7r)$, where $r$ is the squeezing level. The codewords remain orthogonal at all squeezing levels. 
In particular, this code achieves high-precision error correction for both single-photon loss and dephasing, even at moderate squeezing levels. 
Building on this code, we develop quantum error correction schemes that exceed the break--even point, supported by analytical derivations of all necessary quantum gates.
 Our code offers a competitive alternative to previous encodings for quantum computation using continuous bosonic qubits.
\end{abstract}

\maketitle
{\it Introduction.}---Quantum states are fragile due to their high susceptibility to environmental noise, which poses a significant challenge to realizing quantum computation~\cite{Chiaverini2004Dec,Schindler2011May,Mabuchi1996Apr,BibEntry2023Feb,BibEntry2024Dec,BibEntry2013Sep,Gaitan2018Oct}.
Quantum error correction (QEC), which restores quantum information degraded by noise channels through syndrome measurements or reservoir engineering, is therefore essential for fault-tolerant quantum computing~\cite{Kerckhoff2010Jul,Terhal2015Apr,Kosut2008Jan,Barnes2000Jul,Sarovar2005Jul,Cohen2014Dec,Sarma2013Mar,PhysRevLett.131.050601}. This process requires {encoding a single logical qubit redundantly} in large Hilbert spaces, typically {by employing} a block of multiple {physical qubits} or a single higher-dimensional bosonic mode~\cite{Eickbusch2025Dec,PhysRevX.13.031001,Ryan-Anderson2021Dec,Matsuura2020Sep,Fukui2023Oct,Hillmann2022May}.
{In particular, bosonic codes enable QEC at the level of a single bosonic mode with a well-defined set of dominant error channels~\cite{Gottesman2001Jun,Mirrahimi2016Aug,Michael2016Jul,Albert2018Mar,Fukui2017Nov,Lescanne2020May,Zheng2023Jul}. 
	{ 
		While logical error rates cannot be arbitrarily suppressed in these architectures, 
		concatenating bosonic codes with other error-correcting codes, such as the surface code, provides}
	a promising route toward comprehensive error suppression and large-scale fault-tolerant quantum computation~\cite{Putterman2025Feb,Noh2020Jan,Ruiz2025Jan}.}


Bosonic codes hold promise for quantum information processing and thus have garnered considerable attention~\cite{Cai2021Jan,Terhal2020Jul,Joshi2021Apr,Chou2018Sep}. {In many physical implementations of bosonic modes, single-photon loss and dephasing constitute the dominant noise channels, and bosonic codes are often designed to mitigate these errors~\cite{Leviant2022Sep,Lami2023Jun,Grimsmo2020Mar}.} The more the codewords are distributed across the Fock space (i.e., demanding increased coherence), the more dephasing becomes a significant noise source~\cite{Grimsmo2021Jun,PhysRevA.106.022431,PhysRevLett.134.060601}.
Various bosonic codes are designed to correct different types of errors and can be broadly classified into \textit{continuous} and \textit{discrete} codes.  {As in previous works~\cite{Michael2016Jul,Ni2023Apr}, discrete bosonic codes encode logical states in superpositions over a \textit{finite} set of Fock states, whereas continuous codes, such as cat and Gottesman--Kitaev--Preskill (GKP) codes~\cite{Gottesman2001Jun,Fukui2017Nov,Mirrahimi2016Aug}, are supported on an unbounded set of Fock states.
} 


{Being a superposition of a finite number of Fock states, discrete bosonic codes  exhibit a simple structure especially if they are tailored to correct a small number of errors. Nevertheless, this apparent simplicity 
	does not translate into trivial schemes to prepare and control these states,
	as addressing a few Fock states requires a fine-tuned control of the quantum system. Moreover, continuous bosonic codes, such as the squeezed cat or the GKP code, might offer certain advantages for state preparation or control owing to their structure which relies on displacement and squeezing operations  {that can be readily implemented in various quantum platforms, such as superconducting circuits, optical systems, and trapped ions}~\cite{Campagne-Ibarcq2020Aug,Fluhmann2019Feb,Hastrup2022Apr}.
	Noteworthy, this class of codes typically relies on codewords which are strictly not orthogonal, and therefore one needs to choose parameter regimes in which orthogonality is approximately satisfied. 
	Despite some potential advantages in their structure as compared to their discrete counterparts, {existing continuous bosonic codes are typically not tailored to simultaneously address single-photon loss and dephasing}~\cite{Joshi2021Apr,Terhal2020Jul,PhysRevA.106.022431}.}
For example, squeezed cat codes are limited to correcting either single-photon loss or dephasing errors independently; while squeezed Fock codewords are inherently nonorthogonal and thus exhibit diminished performance, particularly at low squeezing levels~\cite{supplement,PhysRevX.13.021004,Bashmakova2025Sep,PhysRevA.107.032423,Korolev2024Oct,Albano2002Sep}.
Developing strictly orthogonal bosonic codes, capable of simultaneously correcting both photon loss and dephasing at experimentally feasible squeezing levels, is therefore essential for scalable bosonic quantum computing.

\begin{figure*}[htp]
	\includegraphics[scale=0.96]{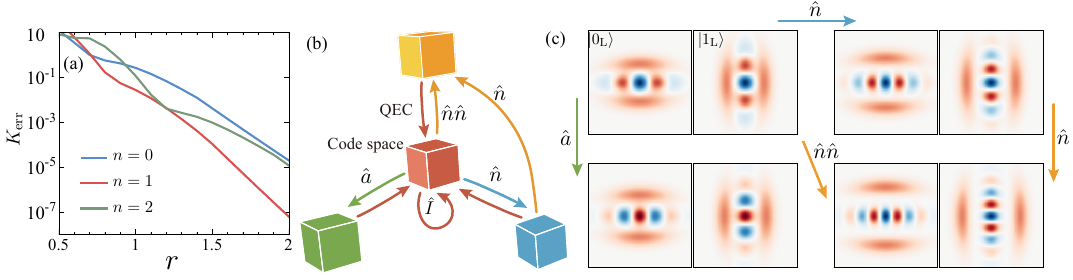}
	\caption{
		{(a)}~Deviation from the KL condition versus the squeezing amplitude $r$ for different values of $n$. An exponential decay is observed, with $n=1$ exhibiting the best performance at large $r$. {(b)}~Errors from the set $[\hat{I}, \hat{a}, \hat{n}, \hat{n}^2]$ acting on the logical subspace define the corresponding error spaces. The code and error spaces approximately satisfy the KL condition $K_{\text{err}}$, enabling recovery via QEC. \textbf{(c)}~ Wigner functions of the codewords  (related by a $\pi/2$ rotation) and associated error states  (also related by a $\pi/2$ rotation)
		at 8~dB squeezing ($r \approx 0.921$).
		{As the $\hat{n}$ and $\hat{n}^2$ operators do not affect the parity of the codewords (as compared to $\hat{a}$), the resulting error states have nonzero overlap with the codewords and with each other. This, in turn, is the reason for the nonvanishing deviation of the KL condition [cf. Eq.~\eqref{seri}].\vspace{-0.36cm}
		}
	}
	\label{Fig1}
\end{figure*}

In this Letter, we combine the advantages of continuous and discrete bosonic codes to construct a  strictly orthogonal code that \textit{achieves  high-precision correction of both single-photon loss and dephasing noise}. The proposed codewords are composed of tailored superpositions of squeezed Fock states, 
where the specific choice of superposition {by construction} ensures exact orthogonality of the codewords---at any squeezing level \( r \). Meanwhile, the squeezing operation distributes the codewords continuously over the infinite-dimensional Fock space. Our analytical results demonstrate an exponential suppression of the infidelity scaling as \( \exp(-7r) \) for combined photon loss and dephasing errors, outperforming all existing continuous bosonic codes.
Therefore, the proposed code enables high-precision correction of both single-photon loss and dephasing errors at moderate squeezing levels. In addition, it facilitates straightforward implementation of logical qubit gates with experimentally accessible operations.
{Specifically, the code supports a simple, error-transparent logical Pauli-$X$ gate that remains effective even within the error space. This property constitutes one of the main advantages of our code.}
Moreover, we analytically design effective quantum gates to implement both autonomous and parity-measurement-based quantum error correction protocols.

{\it Codewords.}---We propose a novel continuous bosonic code that synthesizes discrete and continuous characteristics, specifically designed to correct single-photon loss and dephasing error channels in combination. The logical codewords are defined as
\\
[-16pt]
\begin{equation}
	\begin{split}
		&\left\vert 0_L \right \rangle=\hat{S}\left(r\right)\left(\alpha\left\vert n+2\right\rangle-\beta\left\vert n\right\rangle\right)\\
		&\left\vert 1_L \right\rangle=\hat{S}\left(-r\right)\left(\alpha\left\vert n+2\right\rangle+\beta\left\vert n\right\rangle\right)
	\end{split}
	\label{code}
\end{equation}
\\
[-8pt]
where $\alpha$ ($\vert\beta\vert^2+\vert\alpha\vert^2=1$) is determined by enforcing the orthogonality condition $\langle 0_{\text{L}}\vert 1_{\text{L}}\rangle=0$, $\hat{S}(r)$ denotes the squeezing operator with squeezing amplitude $r$, and $\vert n\rangle$ are the Fock states {($n$ will be specified below)}. 

To evaluate the QEC capability against single-photon loss and dephasing, we examine the Knill-Laflamme (KL) criterion for the error operator set \(\mathbf{E} = \left\{\hat{I}, \hat{a}, \hat{n}, \hat{n}^2\right\}\)~\cite{supplement,Korolev2024Oct,PhysRevA.107.032423}, where $\hat{a}$ and $\hat{n}$ are the annihilation and number operator, respectively.
The KL criterion reads \(\langle u_L | \hat{E}^{\dagger}_i \hat{E}_j | v_L \rangle = C_{ij} \delta_{uv}\)~\cite{PhysRevLett.84.2525}, where the \(C_{ij}\) form a Hermitian matrix, \(\hat{E}_i, \hat{E}_j \in \mathbf{E}\) denote error operators, and \(|u_L\rangle, |v_L\rangle \in \{|0_L\rangle, |1_L\rangle\}\) represent the codewords. 
Exact satisfaction of the KL condition is a prerequisite for complete error correction, and the degree of approximate fulfillment determines the maximal achievable fidelity~\cite{Girvin2023Jun,PhysRevA.56.2567,PhysRevA.55.900,Faist2020Oct}. 
We quantify the deviation from the KL condition using the indicator
\\
[-12pt]
\begin{equation}
	K_{\text{err}} = \sum_{ij} \left| M_{ij}^{00} - M_{ij}^{11} \right|^2 + \left| M_{ij}^{01} \right|^2,
	\label{kl_condition}\vspace{-0.26cm}
\end{equation}
\\
[-2pt]
where \( M_{ij}^{\mu\nu} = \langle \mu_{\text{L}} | \hat{E}_i^\dagger \hat{E}_j | \nu_{\text{L}} \rangle \).
A smaller value of \( K_{\text{err}} \) implies a closer approximation to the ideal KL condition and hence stronger error-correcting performance.

The condition \( M^{00}_{ij} = M^{11}_{ij} \) is  satisfied for the code in Eq.~(\ref{code}), and hence the first term in Eq.~(\ref{kl_condition}) vanishes identically. This indicates that the error process does not distinguish between logical basis states—a distinct advantage over the squeezed cat code. {Consequently, only the off-diagonal terms quantifying the coherence between logical states 
	affected by errors
	contribute to the deviation from the KL condition.}
Although some off-diagonal terms \( M^{01}_{ij} \) are nonzero, they remain exponentially small, scaling as \( \sim e^{-7r} \) for odd \( n \) and \( \sim e^{-5r} \) for even \( n \). This difference arises from the distinct probability amplitude distributions of squeezed even and odd Fock states in Fock space. Since increasing \( n \) does not improve the scaling behavior but increases the average excitation number—which is approximately proportional to the error occurrence probability—we select \( n = 1 \) as the optimal choice and focus on it hereafter.

We simulate \( K_{\text{err}} \) in Fig.~\ref{Fig1}(a) for different values of $n$, confirming that \( n = 1 \) is the optimal choice and that the QEC performance improves with increasing squeezing. The nonzero terms responsible for the deviation from the KL condition can be approximately expanded in exponentials of the squeezing amplitude \( r \): 
\\
[-12pt]
\begin{equation}
	\langle1_{L}\vert\hat{n}^{i}\vert0_{L}\rangle=S_ie^{-7r}+\mathcal{O}(e^{-9r}), \;\;\; i=1, 2, 3,4 \label{seri}
\end{equation}
\\
[-12pt]
with $S_i\in \mathbf{S}=\{\pm6.4\sqrt3,\;3.2\sqrt2(5\mp\sqrt6),\;
24\sqrt2\mp36.8\sqrt3,\;
8\sqrt2(31\pm5\sqrt6)\}$,
where $\pm$ correspond to different solutions of $\alpha$. The solutions for the coefficient $\alpha$ are not unique; however, this ambiguity does not affect the exponential scaling. 
{Note that all other terms of the KL conditions are strictly satisfied. 
	As we will detail next, the scaling of the off-diagonal terms in \( K_{\text{err}} \) achieved by our proposed code represents a significant improvement over previous continuous bosonic codes.}

The value of \( K_{\text{err}} \) reaches the order of \( 10^{-2} \) at \( r \approx 0.921 \) (approximately 8~dB), a squeezing level readily achievable in current experiments~\cite{PhysRevX.13.021004}. For the error set \( \mathbf{E} \), this corresponds to a reduction in deviation by more than \textit{three} and \textit{two} orders of magnitude compared to the squeezed cat and squeezed Fock codes, respectively.
This performance gap stems from the substantial overlap of squeezed cat codewords under the combined action of \( \hat{a} \) and \( \hat{n}^m \). Consequently, the code performs significantly worse for error sets that simultaneously include both operators, such as \(\mathbf{E}\), even though subsets like \( \{ \hat{I}, \hat{a} \} \) or \( \{ \hat{I}, \hat{n}, \hat{n}^2 \} \) remain approximately correctable.
Similarly, the squeezed Fock code suffers from inherent non-orthogonality and off-diagonal terms \( M^{01}_{ij} \) that decay no faster than \( \sim e^{-3r} \), resulting in markedly inferior error suppression relative to our code.

Short-time quantum dynamics gives rise to three distinct error subspaces, whose relation to the code space is depicted in Fig.~\ref{Fig1}(b). The corresponding Wigner functions of the codewords related by a \( \pi/2 \) rotation and their error-transformed states, are shown in Fig.~\ref{Fig1}(c).
Consequently, the logical Pauli \( X \) operator is given by \(\hat{X}_L = \exp(-i \frac{\pi}{2} \hat{n})\), which remains effective within the error spaces; hence, it is an error-transparent gate. 
{Furthermore, the Pauli-\( Z \) operator can be implemented as \( \hat{Z}_L = \exp[-i \hat{H}_z(\hat{a}, \hat{a}^\dagger)] \), where \( \hat{H}_z \) is expressed as a power-series expansion in the operators \( \hat{a} \) and \( \hat{a}^\dagger \)~\cite{supplement}.
} 
{Finally, the algebraic properties of the Pauli operators and the structure of the codewords allow for a simplified preparation scheme—once one codeword is prepared, the other can be obtained through a  rotational operation.}




\begin{figure}
\includegraphics[scale=0.92]{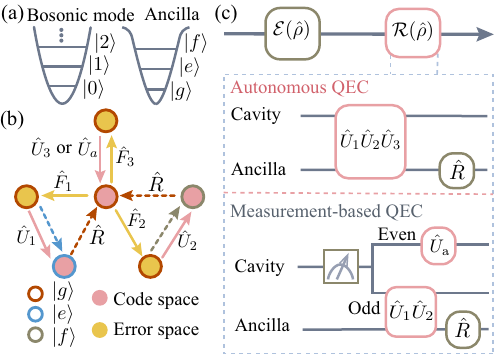}
\vspace{-0.22cm}
\caption{
	{(a)}~The encoded bosonic mode resides in an infinite-dimensional Hilbert space, while the auxiliary system is a discrete-level system, such as a qutrit. {(b)}~The bosonic mode is initialized in an encoded state $|\psi_{\mathrm{L}}\rangle$, and the auxiliary system in its ground state \( |g\rangle \). Noise operators \( \hat{F}_i \), derived from the error set \( \mathbf{E} \), map the code space into approximately orthogonal error subspaces. {Control operations \( \hat{U}_i \) conditionally restore the encoded state while exciting the auxiliary system. A reset operation \( \hat{R} \) returns the auxiliary system to \( |g\rangle \), completing a QEC cycle.} {(c)}~{After the error channel $\mathcal{E}(\cdot)$, two alternative recovery schemes $\mathcal{R}(\cdot)$ are considered. 
		The stroboscopic autonomous QEC completes the recovery without measurements, 
		while the measurement-based protocol uses a parity measurement to identify the error: 
		even parity invokes $\hat{U}_a$ [Eq.~(S23)] for direct recovery, 
		odd parity triggers a two-step correction via $\hat{U}_1$ and $\hat{U}_2$. } Iterating these cycles enables long-term protection of encoded quantum information. \vspace{-0.26cm}}
\label{Fig2}
\end{figure}

{\it Error correction schemes.}--- 
The strong alignment of the code with the KL conditions, together with the odd parity structure of the code space, enables the systematic design of QEC schemes {by using an ancilla three-level system, as shown in Fig.~\ref{Fig2}(a)}, including both stroboscopic autonomous and parity-measurement-based approaches. {Note that stroboscopic autonomous QEC proceeds via deterministically  triggered recovery operations, unlike continuous autonomous QEC based on engineered dissipation}. The encoded bosonic mode first undergoes a noise channel $\mathcal{E}(\cdot)$, followed by a recovery operation $\mathcal{R}(\cdot)$. For a short evolution time $\tau$, the combined effect of single-photon loss and dephasing is described by a Kraus expansion,
$	\mathcal{E}[\hat{\rho}] \approx \sum^3_{i=1} \hat{A}_i \hat{\rho} \hat{A}_i^\dagger,
$
where the Kraus operators are given by
\( \hat{A}_1 \approx \hat{I} - \frac{\kappa \tau}{2} \hat{n} - \frac{\kappa_\phi \tau}{2} \hat{n}^2 \), 
\( \hat{A}_2 \approx \sqrt{\kappa_\phi \tau}\, \hat{n} \), and \( \hat{A}_3 \approx \sqrt{\kappa \tau}\, \hat{a} \)  
~\cite{supplement}.
Here, \( \kappa \) and \( \kappa_\phi \) are the single-photon loss and dephasing rates, respectively.
To construct the QEC channel, we diagonalize the symmetric matrix  
\( \boldsymbol{J} = V \boldsymbol{\Lambda} V^\dagger \),  
with elements \( J_{ij} = \langle u_L | \hat{A}_i^\dagger \hat{A}_j | u_L \rangle \). 
The noise channel can be recast as  
\(
\mathcal{E}[\hat{\rho}] \approx \sum_i \hat{F}_i \hat{\rho} \hat{F}_i^\dagger,
\)  
where the transformed Kraus operators are~\cite{Girvin2023Jun}  
\\
[-18pt]
\begin{equation}
\hat{F}_i = \sum_{k=1}^3 V_{ki} \hat{A}_k, \;\;\; i=1,2,3
\vspace{-0.2cm}
\end{equation} 
\\
[-6pt]
These operators satisfy  
\( \hat{P}_L \hat{F}_i^\dagger \hat{F}_j \hat{P}_L \approx \Lambda_{ij} \delta_{ij} \hat{P}_L \),  
where \( \hat{P}_L \) denotes the projector onto the code space.  
This ensures that the \( \hat{F}_i \) operators map the code space onto orthogonal error subspaces, enabling efficient recovery.

As shown in Fig.~\ref{Fig2}(a,b), we couple the encoded bosonic mode to an ancillary qutrit initialized in its ground state \( \vert g \rangle \) to identify and correct the errors \( \hat{F}_i \), resulting in the initial joint state \( \vert \psi_{\text{L}}, g \rangle \). A sequence of 
{ unitary operations}
\( \hat{U}_i \) [\( i = 1,2,3\) in Eq.~(\ref{eq:unitaries}) and $ i=a$ in Eq.~(S23)] is then applied to coherently restore the system to the code space. {Since only $\hat{a}$ flips the parity of the encoded state, the respective error syndrome can be extracted via a parity measurement: based on the measurement outcome, we apply $\hat{U}_a$ for the parity-flip case or the common recovery $\hat{U}_2\hat{U}_1$ otherwise, followed by the Hermitian operation $\hat{R}$ that coherently returns the qutrit to the ground state $\vert g\rangle$ without inducing excitations, as shown in Fig.~\ref{Fig2}(c).}
While conceptually similar to the autonomous scheme, the measurement-based approach requires active readout and feedforward, introducing additional overhead in experimental implementations. We therefore focus on the stroboscopic autonomous scheme in the main text, with the measurement-based protocol detailed in ~\cite{supplement}.

The recovery 
unitary operations $\hat{U}_i$ correct errors associated with the operators $\hat{F}_i$ while acting as the identity operator in the other orthogonal error subspaces. As a result, they do not interfere with errors arising from other Kraus operators. Following the recovery process, the encoded state $\vert \psi_L \rangle$ is approximately restored, while the ancilla qutrit transitions to different states depending on the specific error that occurred~\cite{supplement}. Finally, the ancilla qutrit is rapidly reset to its ground state via a strongly dissipative interaction $R$ with a reservoir, completing the QEC cycle without measurement and thereby enabling  autonomous QEC.
The entire QEC cycle can be described by the following equation
\\
[-12pt]
\begin{equation}
\mathcal{R}_{\text{a}}\circ\mathcal{E}[\hat{\rho}_L]
= R\Big[  \hat{U} \, \mathcal{E}[\hat{\rho}_{\text{L}}] \otimes \vert g\rangle\langle g\vert \, \hat{U}^\dagger \,  \Big] \\
\approx \hat{\rho}_L \otimes \vert g\rangle\langle g\vert,
\label{sequation_10}
\end{equation}
\\
[-12pt]
where $\mathcal{R}_{\text{a}}$ is the autonomous QEC channel, $\hat{U}=\hat{U}_3\hat{U}_2\hat{U}_1$, $\hat{\rho}_L$ is the encoded state $\hat{\rho}_L=\vert\psi_{L}\rangle\langle\psi_L\vert$, and $R$ stands for the reset of the ancilla qutrit to its ground state.

{Here, we present an 
analytical approach for the above  recovery unitary operations
by incorporating an ancilla qutrit system, with the detailed expressions given by}
\\
[-18pt]
\begin{equation}
\begin{split}
	\hat{U}_1&=\hat{L}_1\vert e\rangle\langle g\vert+\hat{L}^{\dagger}_1\vert g\rangle\langle e\vert+\hat{U}_{1,\text{re}}, \\
	\hat{U}_2&=\hat{L}_2\vert f\rangle\langle g\vert+\hat{L}^{\dagger}_2\vert g\rangle\langle f\vert+\hat{U}_{2,\text{re}},\\
	\hat{U}_3&=\left(\hat{L}_3+\hat{L}^{\dagger}_3+\hat{I}-\hat{P}_{L}-\hat{P}_{F_3}\right)\vert g\rangle\langle g\vert+\hat{U}_{3,\text{re}},		
\end{split}	\label{eq:unitaries} 
\end{equation}
\\
[-12pt]
where $\vert e\rangle$, $\vert f\rangle$ are two excited states 
{ used to discriminate between different types of errors,}
the operator $\hat{L}_i=\vert0_{L}\rangle\langle0_{F_{i}}\vert+\vert1_{L}\rangle\langle1_{F_{i}}\vert$ is designed to recover information from the error space into the code space ($\vert u_{F_i}\rangle=\hat{F}_{i}\vert u_L\rangle/\|\hat{F}_{i}\vert u_L\rangle\|$), $\hat{P}_{F_i}$ is the projector operator of the $i$th error space,  and $\hat{U}_{i,\text{re}}$ supplements  
$\hat{U}_i$ to ensure it forms a unitary operator and leave the recovered information unaffected ($\hat{U}_{i,\text{re}}$ is provided in~\cite{supplement}).  These 
unitary operations
correct the corresponding errors without affecting other parts of the error space or states corrected after other errors occur. 
The availability of analytic expressions 
for the recovery unitaries~\eqref{eq:unitaries} allows for their efficient implementation using well-established quantum control techniques, such as gradient ascent pulse engineering (GRAPE) and machine-learning-based optimization~\cite{Khaneja2005Feb,deFouquieres2011Oct,Johansson2012Aug,Mercurio2025Sep,Johansson2013Apr,Lambert2024Dec,Eickbusch2022Dec}.
Alternatively,
we could use two-qubit or multilevel ancilla systems to design these 
unitary operations
with the construction methodology similar to that of the qutrit ancilla system~\cite{supplement}. {In principle, arbitrary unitary operations on the system can be achieved through optimal control techniques. As an example, we consider a superconducting cavity coupled to a transmon, where the composite gate $\hat{U}$ is realized by controlling the quadratures $\hat{x}$ and $\hat{p}$ of the bosonic mode, achieving a fidelity exceeding $0.99$~\cite{supplement}. The control fields can be flexibly adjusted to enhance the performance further, but such refinements are beyond the scope of this work.

}

\begin{figure}
\includegraphics[scale=0.76]{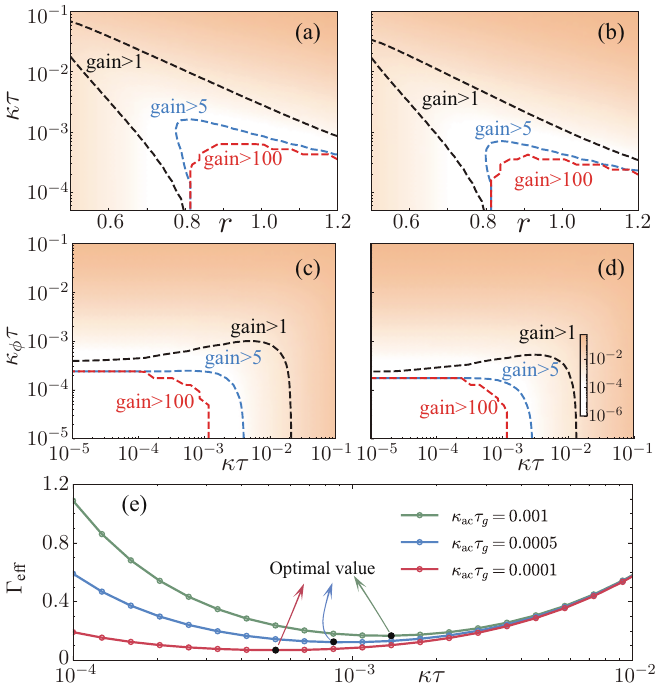}
\caption{
	{Panels (a) and (b) show the channel {infidelity} per QEC cycle versus the dimensionless time scale $\kappa\tau$ and squeezing amplitude $r$ for photon-loss--to--dephasing rate ratios $\kappa/\kappa_\phi \approx 5.5$ and $2.5$, respectively.
		Panels (c) and (d) show the channel infidelity versus the single-photon loss and dephasing rates for $r=0.9$ and $r=1.0$. (e) Effective decay rate $\Gamma_{\mathrm{eff}}$ of the channel {fidelity} versus the idling time $\kappa \tau$ for different gate-time scales $\kappa_{ac}\tau_g$ at $r=0.9$.
		All results are obtained by simulating the full master equation. \vspace{-0.56cm}
}}
\label{Fig3}
\end{figure}
{Figure~\ref{Fig3} shows the average infidelity versus the squeezing
amplitude $r$ and the dimensionless time scale $\kappa\tau$, where $\tau$
is the idle time and the fidelity is
averaged over the six Pauli eigenstates~\cite{PhysRevLett.125.260509}. For $r\gtrsim0.8$ (corresponding to $\gtrsim6.95$~dB of squeezing), the
autonomous QEC protocol can yield a gain
{$(1-\bar{F}_{\mathrm{be}})/(1-\bar{F}_{\mathrm{auto}})>100$ 
	for $\kappa/\kappa_{\phi} \approx 5.5$ and $\kappa/\kappa_{\phi} \approx 2.5$, as shown in Fig.~\ref{Fig3}(a) and (b) (respectively).} 
Here, $\bar{F}_{\mathrm{be}}$ denotes the channel fidelity of a qubit encoded in the
bare Fock states $\vert 0\rangle$ and $\vert 1\rangle$ without QEC, while $\bar{F}_{\mathrm{auto}}$ corresponds to the autonomous QEC channel fidelity 
{under the assumption of instantaneous recovery operations $\hat{U}_i$~\eqref{eq:unitaries}.}
{Our results show that moderate squeezing is required for a gain $>1$, as the KL enhancement must outweigh the increased the average photon number of the codewords. Notably, for amplitude $r$ slightly above 0.8, the gain exceeds 1 for arbitrarily small $\kappa\tau$, as the KL contribution dominates over the excitations.}
{Importantly, we simulate the full noise channel corresponding to photon-loss and dephasing noise. Only in the small time limit ($\kappa \tau, \kappa_\phi \tau \ll 1$), 
	this channel can be approximated by the error set $\mathbf{E} = \{\hat{I},\hat{a},\hat{n},\hat{n}^2\}$. 
	Beyond this limit, the channel can induce cumulative error processes that are not accounted for by our code.}
Within the explored parameter regime, the
optimal performance occurs around $r=0.9$, which remains effective over a
broad range of $\kappa/\kappa_{\phi}$ and outperforms stronger squeezing
($r=1$), as shown in Fig.~\ref{Fig3}(c,d). Using a near-optimal recovery channel, such as the Petz recovery map~\cite{Kwon2022Jan,Li2025May}, efficient error correction persists down to $r=0.6$, with a gain exceeding $10$ at $r=0.6$ and $\kappa\tau = \kappa_{\phi}\tau\approx 0.01$~\cite{supplement}. At the corresponding squeezing levels and time scales, the resulting infidelity is lower than that of both the squeezed cat code and the GKP code~\cite{PhysRevA.106.022431,Grimsmo2021Jun,PhysRevLett.125.260509}.

{Finally, we assess ancilla errors on the channel fidelity by implementing the operations $\hat{U}_i$ using gates of finite duration.}
We further analyze the effect of ancilla decay $\kappa_{\mathrm{ac}}$ by extracting the effective
decay rate of the channel fidelity,
$\Gamma_{\mathrm{eff}}\propto
-\ln[\bar{F}_{\mathrm{auto}}(\tau_{\mathrm{al}})]/\tau_{\mathrm{al}}$, as shown
in Fig.~\ref{Fig3}(e), where  $\tau_{\mathrm{al}}=\tau+\tau_g$ is the duration of a single QEC cycle. {Here, we model the ancilla as a decaying mode $\hat{b}$ and incorporate the unitary process $\hat{U}_3 \hat{U}_2 \hat{U}_1 = \exp(-\hat{H}_{\mathrm{eff}} \tau_g)$ into the system dynamics with the effective Hamiltonian $\hat{H}_{\mathrm{eff}}$.} 
For a fixed squeezing amplitude $r$, shorter gate times $\tau_g$ result in improved QEC performance. Moreover, ancilla dissipation leads to the
emergence of an optimal idle time. For $r=0.9$, the optimal idle time is on
the order of $10^{-3}$, consistent with the experimentally optimal time for
implementing GKP encoding ($\kappa\tau\approx0.00074$), and the corresponding gate times fall
within experimentally achievable regimes~\cite{Sivak2023Apr}. {For these optimal idle-time scales, the gain exceeds 5, comparable to the case where the noise of the ancillary qubit is neglected}. Requiring
significantly less squeezing while supporting realistic idling times
$\tau$, our code achieves high QEC performance and is compatible with a
wide range of experimental platforms, including trapped ions, optical
systems, and 3D superconducting microwave
cavities~\cite{deNeeve2022Mar,Brock2025May,Cai2022Jun,Wang2016May,
	PhysRevLett.132.230602,Sivak2023Apr,Hillmann2020Oct,Rosenblum2018Feb,
	Rosenblum2018Jul,Reinhold2020Aug}.

} 

{
{We emphasize that binomial codes can be designed to satisfy the KL conditions for the same error set as our proposed code. 
	The result is a high-order code involving superpositions up to Fock state $\vert 10 \rangle$. Whereas these codes 
	offer an error correction performance comparable to ours, they are experimentally challenging due to their reliance on multiple high-Fock components~\cite{Michael2016Jul}.}
So far, experiments have only realized the lowest-order binomial code with superpositions up to Fock state $\vert 4 \rangle$, which are limited to correcting single-photon loss~\cite{Hu2019May}. In contrast, encodings within low-Fock subspaces such as Fock states $\vert 1 \rangle$ and $\vert 3 \rangle$ are well established experimentally~\cite{Wang2022Jun,Sayrin2011Sep,Krastanov2021Jan}. Our code builds on this subspace, requiring only an additional squeezing operation for codeword preparation. This results in a preparation process that is experimentally more accessible—even simpler than that of the lowest-order binomial codes—while offering enhanced error correction. Consequently, our code combines improved performance with high experimental feasibility.
}


{\it Conclusion.}---We introduced a squeezed bosonic code that robustly corrects for both single-photon loss and dephasing in continuous-variable quantum systems under experimentally feasible conditions. Exploiting the infinite-dimensional Hilbert space of bosonic modes, our code features error-correcting capabilities that scale exponentially with the squeezing amplitude [$\propto \exp(-7r)$] while maintaining orthogonal codewords at all squeezing levels. {As a result, the QEC performance improves exponentially with the squeezing for the error set $\mathbf{E}$, enabling high QEC efficiency at relatively low squeezing. Moreover, the code applies equally when $\hat{a}$ is replaced by $\hat{a}^\dagger$, which satisfies the same error-correction conditions.
} 
Building on this framework, we developed measurement-based and autonomous QEC protocols and provide an analytical description of the required 
recovery operations.
Our analysis establishes that these QEC schemes can be implemented in a bosonic mode coupled to a qutrit. In conclusion, this new code represents a substantial advancement over previous continuous-variable bosonic qubit encodings for quantum computation.


{\it Acknowledgments}---
F.N. is supported in part by the Japan Science and Technology Agency (JST) 
[via the CREST Quantum Frontiers program Grant No. JPMJCR24I2, 
the Quantum Leap Flagship Program (Q-LEAP), the Moonshot R\&D Grant Number JPMJMS256E, 
and the ASPIRE program (Grant Number JPMJAP2513)],
and the Office of Naval Research (ONR) Global (via Grant No. N62909-23-1-2074).
  C.G. was partly supported by
a RIKEN Incentive Research Project Grant.

\textit{Data} availability—The data supporting the findings of this study are available from the author upon reasonable request.




\bibliography{main}

\end{document}


	\title{Supplementary Materials for ``Quantum Error Correction with Superpositions of Squeezed Fock States''}
	
\author{Yexiong Zeng~\orcid{0000-0001-9165-3995}}

\affiliation{RIKEN Center for Quantum Computing, RIKEN, Wakoshi, Saitama 351-0198, Japan}
\affiliation{Key Laboratory of Low-Dimensional Quantum Structures and Quantum Control of Ministry of Education,
	Department of Physics and Synergetic Innovation Center for Quantum Effects and Applications,
	Hunan Normal University, Changsha 410081, China}

\author{Fernando Quijandría~\orcid{0000-0002-2355-0449}}\affiliation{RIKEN Center for Quantum Computing, RIKEN, Wakoshi, Saitama 351-0198, Japan}	

\author{Clemens Gneiting~\orcid{0000-0001-9686-9277}}
\altaffiliation[clemens.gneiting@riken.jp]{}
\affiliation{RIKEN Center for Quantum Computing, RIKEN, Wakoshi, Saitama 351-0198, Japan}

\author{Franco Nori~\orcid{0000-0003-3682-7432}}
\altaffiliation[fnori@riken.jp]{}
\affiliation{RIKEN Center for Quantum Computing, RIKEN, Wakoshi, Saitama 351-0198, Japan}
\affiliation{Quantum Research Institute and Department of Physics, University of Michigan, Ann Arbor, Michigan, 48109-1040, USA}

	

	

	\maketitle
	\tableofcontents
	
\section{Codeword Design}
We consider superposition states composed of squeezed Fock states as our codewords, explicitly defined as
\begin{equation}
	\label{codewords}
	\begin{split}
		\vert 0_{\text{L}} \rangle 
		&= \hat{S}(r)\big(\alpha \vert n+2 \rangle 
		- \sqrt{1-\alpha^2}\, \vert n \rangle \big)  = \alpha \vert n+2, r \rangle 
		- \sqrt{1-\alpha^2}\, \vert n, r \rangle , \\
		\vert 1_{\text{L}} \rangle 
		&= \hat{S}(-r)\big(\alpha \vert n+2 \rangle 
		+ \sqrt{1-\alpha^2}\, \vert n \rangle \big)  = \alpha \vert n+2, -r \rangle 
		+ \sqrt{1-\alpha^2}\, \vert n, -r \rangle .
	\end{split}
\end{equation}
{where \( \vert n \rangle \) denotes a Fock state, and the squeezed number state is defined as \( \vert n, r \rangle = \hat{S}(r) \vert n \rangle \), where $\hat{S}(r) = \exp[r(\hat{a}^{2} - \hat{a}^{\dagger 2})/2]$ is the squeezing operator with $r$ a real parameter quantifying the squeezing level.
Finally,
\( \alpha \) 
is a real parameter determined by the orthogonality condition \( \langle 0_{\text{L}} \vert 1_{\text{L}} \rangle = 0 \). 
{We can give the two solutions of $\alpha$ explicitly,\begin{equation}
\alpha_{1/2}=\sqrt{\frac{2\left[2 \cosh(2r)^4 + 2\cosh(2r)^2 + 
	3\sinh(2r)^2 \cosh(2 r)^2 \mp 
	2\sqrt{6}\sinh(2r)\cosh(2r)^2\right]}{9\sinh(2r)^4 - 12\sinh(2r)^2 + 4 \cosh(2r)^4 + \
	8\cosh(2r)^2 + 12\sinh(2r)^2\cosh(2r)^2 + 4}}.
\label{alpha1}
\end{equation}}
}
The Fock-space representation of squeezed Fock states differs markedly between even and odd \( n \). The corresponding occupation probability is given by
\begin{equation}
	\left\langle n|m,r\right\rangle\underline{{{}}}=\frac{\left(m!n!\right)^{\frac{1}{2}}}{\mathrm{cosh}(r)^{\frac{n+m+1}{2}}}\times\sum_{k}^{\mathrm{min}(m,n)}\left[\frac{\sinh(r)}{2}\right]^{\frac{n+m-2k}{2}}\frac{(-1)^{\frac{n-k}{2}}}{k!\left(\frac{m-k}{2}\right)!\left(\frac{n-k}{2}\right)!},
\end{equation}
where $k$ takes even or odd values in correspondence with the even or odd nature of $m$ and $n$, respectively 
\cite{Nieto1997May}.
From the above expression, in the limit \( r \to \infty \), it can be shown that the overlap \( \langle n | m, r \rangle \) scales as \( \cosh(r)^{-\frac{3}{2}} \) for odd \( n \), and as \( \cosh(r)^{-\frac{1}{2}} \) for even \( n \).

\begin{figure}[ht]
	\centering
	\includegraphics[width=1.01\textwidth]{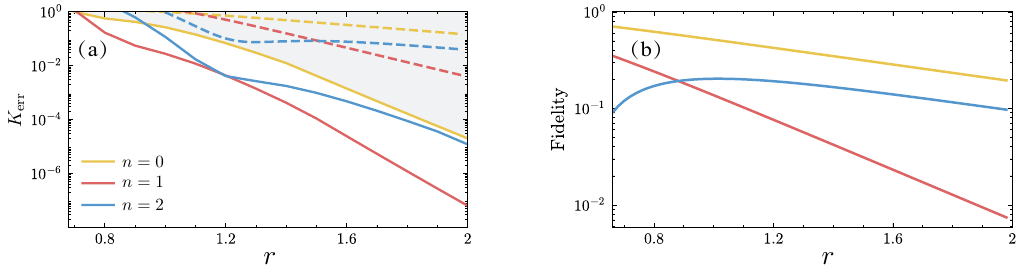}
	\caption{(a) The {deviation $K_{\text{err}}$} of our code (solid lines) and the squeezed Fock code (dashed lines) with respect to the squeezing parameter \(r\) is presented for different values of \( n \). {The shaded region above the continuous yellow curve ($n=0$) indicates the range of \( K_{\text{err}} \) values for the squeezed Fock code, while our code remains well below this range for all squeezing amplitudes \( r \).}  
		(b) The overlap, captured by the fidelity, between the {codewords of the} squeezed Fock code is shown to remain high even under strong squeezing. In contrast, our code is orthogonal by construction.}
	\label{fig:lru_scheme}
\end{figure} 
{In most physical implementations of bosonic modes, single-photon dissipation and dephasing dominate over other error mechanisms. We therefore model the system dynamics using the master equation}
\begin{equation}
	\frac{d\hat{\rho}}{dt}=\frac{\kappa}{2}\mathcal{D}[\hat{a}]\hat{\rho} +\frac{\kappa_{\phi}}{2}\mathcal{D}[\hat{n}]\hat{\rho}
	\label{master_equation}
\end{equation}
where $\kappa$ ($\kappa_{\phi}$) is the single photon loss (dephasing) rate, and $\mathcal{D}[\hat{x}]=2\hat{x}\hat{\rho}\hat{x}^{\dagger}-\hat{x}^{\dagger}\hat{x}\hat{\rho}-\hat{\rho}\hat{x}^{\dagger}\hat{x}$ is the Lindblad super-operator.

{
The short-time non-unitary dynamics can be expressed in the Kraus form.
The Kraus operators for photon loss and dephasing over a short interval $\tau$ are
$\hat{A}_2 = \sqrt{\kappa_{\phi} \tau}\,\hat{a}^{\dagger}\hat{a}$ and $\hat{A}_3 = \sqrt{\kappa \tau}\,\hat{a}$, respectively. 
For an infinitesimal time step, we consider at most a single quantum jump from either $\hat{A}_2$ or $\hat{A}_3$, with the remaining operator describing the non-Hermitian evolution in the absence of quantum jumps:
\begin{equation}
	\hat{A}_1 =\exp\Big(-\frac{\kappa \tau}{2} \hat{n} - \frac{\kappa_\phi \tau}{2} \hat{n}^2\Big) 
\approx \hat{I} - \frac{\kappa \tau}{2} \hat{n} - \frac{\kappa_\phi \tau}{2} \hat{n}^2\approx \sqrt{\hat{I} - \hat{A}_2^\dagger \hat{A}_2 - \hat{A}_3^\dagger \hat{A}_3},
\end{equation}
where we have neglected terms of higher order in $\tau$. A similar treatment can be found in Refs.~\cite{10.21468/SciPostPhysLectNotes.70,BibEntry2023Dec}.
}Thus, we recast the master equation in terms of Kraus operators at short time scales $\tau\ll 1$
\begin{equation}
\hat{\rho}(t+\tau)=\mathcal{E}[\hat{\rho}(t)]\approx\sum_i\hat{A}_i\hat{\rho}(t)\hat{A}^{\dagger}_i
\label{kraus_oper}.
\end{equation} 

Consequently, the error set accounting for single-photon loss and dephasing can be represented as \( \mathbf{E} = \{ \hat{I},\, \hat{a},\, \hat{n},\, \hat{n}^2 \} \). If the elements of this set satisfy the Knill-Laflamme (KL) condition, the associated errors are fully correctable. The KL condition reads \( \langle u_{\text{L}} \vert \hat{A}_i^\dagger \hat{A}_j \vert v_{\text{L}} \rangle = C_{ij} \delta_{uv} \), {where $C$ is a Hermitian matrix.} 
{Given the structure of the codewords~(\ref{codewords}), it is straightforward to verify that \( \langle 0_{\text{L}} \vert \hat{n}^m \vert 0_{\text{L}} \rangle = \langle 1_{\text{L}} \vert \hat{n}^m \vert 1_{\text{L}} \rangle \).
Therefore, the two logical basis states exhibit identical error probabilities under the same noise processes. 
In addition, the codewords always satisfy \( \langle u_{\text{L}} \vert \hat{a} \hat{n}^m \vert v_{\text{L}} \rangle = 0 \).}
Furthermore, the KL conditions can be systematically expanded in powers of \( e^{-r} \) in the large-squeezing limit \( r \to \infty \). For odd \( n \), we find 
\begin{equation}
\langle 1_{\text{L}} \vert \hat{n}^m \vert 0_{\text{L}} \rangle \propto e^{-7r} + \mathcal{O}(e^{-9r}),
\end{equation}
whereas for even \( n \), the scaling becomes \( e^{-5r} + \mathcal{O}(e^{-7r}) \).

{We evaluate the deviation from the KL condition, denoted as $K_{\text{err}}$ in the main text, for both our proposed code 
and the squeezed Fock code~\cite{BibEntry2023Dec}
\begin{align}\label{eq:sqfc}
    \ket{0_{\rm L}} &= \hat{S}(r) \ket{n}, \nonumber\\
    \ket{1_{\rm L}} &= \hat{S}(-r) \ket{n},
\end{align}
at various excitation numbers $n$, as shown in Fig.~\ref{fig:lru_scheme}(a). Notably, the case $n=1$ not only exhibits low excitation levels, but also yields superior error correction performance. 
{It is evident that our code 
exhibits a smaller deviation from the KL condition compared to the squeezed Fock code
at \textit{all} squeezing levels.}

Furthermore, as illustrated in Fig.~\ref{fig:lru_scheme}(b), the squeezed Fock states are not strictly orthogonal, whereas our codewords 
are orthogonal by design.} 
In the following, we focus on the minimal nontrivial case \( n = 1 \), which corresponds to the lowest-photon-number encoding within this code family while preserving nontrivial error-correcting capability. In this $n=1$ case, the remaining nonvanishing KL conditions admit the following asymptotic expansions:
\begin{align}
	\langle1_{L}\vert\hat{n}\vert0_{L}\rangle&=\pm\frac{32\sqrt{3}e^{-7r}}{5}-\frac{64\sqrt{2}e^{-9r}}{25}+\mathcal{O}(e^{-11r}), \nonumber\\
	\langle1_{L}\vert\hat{n}^2\vert0_{L}\rangle&=\frac{32\sqrt{2}}{25}\left(2\pm35\sqrt{6}\right)e^{-9r}-\frac{16\sqrt{2}}{5}\left(5\pm\sqrt{6}\right)e^{-7r}+\mathcal{O}(e^{-11r}),\nonumber
	\\
	\langle1_{L}\vert\hat{n}^{3}\vert0_{L}\rangle&=\left(24\sqrt{2}\mp\frac{184\sqrt{3}}{5}\right)e^{-7r}-\frac{16\sqrt{2}}{25}\left(502\pm105\sqrt{6}\right)e^{-9r}+\mathcal{O}(e^{-11r}), \label{seri}
	\\
	\langle1_{L}\vert\hat{n}^{4}\vert0_{L}\rangle&=\left(640\sqrt{2}\mp\frac{6944\sqrt{3}}{5}\right)e^{-9r}+8\sqrt{2}\left(31\pm5 \sqrt{6}\right)e^{-7r}+\mathcal{O}(e^{-11r}).\nonumber	
	\label{series}
\end{align}
Evidently, in the limit \( r \to \infty \), the KL condition is exactly satisfied. Notably, our expansion contains only \textit{terms of order 
equal or higher than \( e^{-7r}\)}, indicating that the code provides a good approximation to the KL condition for single-photon loss and dephasing even with moderate squeezing. The error-correcting performance further improves with increasing \( r \). 

{
	\begin{figure}[t]
		\centering
		\includegraphics[width=\textwidth]{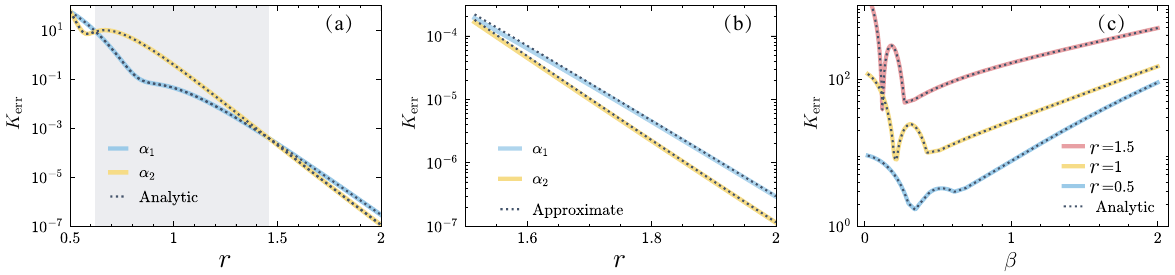}
		\caption{{(a) KL deviation \( K_{\text{err}} \) between the two codewords corresponding to the two solutions of \( \alpha \). {The solid lines show the numerical results for $K_{\mathrm{err}}$ obtained with a Fock-space truncation of 500, while the dotted lines indicate the exact analytical results, which are in perfect agreement with the numerical calculations. {The shaded area indicates the parameter region where $\alpha_1$ performs better, while outside this region $\alpha_2$ is favorable, allowing one to choose the optimal code according to the squeezing amplitude $r$.}}  
				(b) Approximate KL deviation \( K_{\text{err}} \) from Eq.~\ref{seri}, obtained via series expansion (dotted lines) and numerical simulation (solid lines).  
				(c) KL divergence of the squeezed cat code as a function of the amplitude \( \beta \) for various squeezing levels, under the error set \( \{\hat{I}, \hat{a}, \hat{n}, \hat{n}^2\} \) that includes single-photon loss and dephasing.}
		}
		\label{fig:lru_scheme_2d}
	\end{figure}
We compare the numerical and analytical results in Fig.~\ref{fig:lru_scheme_2d}~(a), and find that they are in good agreement. {The two solutions for $\alpha$ in Eq. (\ref{alpha1}), obtained from the orthogonality condition $\langle 0_{\text{L}} | 1_{\text{L}} \rangle = 0$, leave the exponential scaling unchanged, while an appropriate choice of $\alpha$ can improve the QEC performance for a given squeezing amplitude $r$.} Furthermore, we evaluate $K_{\text{err}}$ using the approximate series expansion given in Eq.~\ref{fig:lru_scheme_2d}~(b), which also shows excellent agreement with the exact results, thereby confirming the effectiveness of the approximation solution. }

In contrast, the squeezed Fock code~\eqref{eq:sqfc}
is not inherently orthogonal, with the off-diagonal KL terms scaling as \( \langle 0_{\text{L}} \vert \hat{n}^m \vert 1_{\text{L}} \rangle \propto e^{-3r} \)~\cite{BibEntry2023Dec}. As a result, achieving approximate orthogonality requires significantly larger squeezing. For instance, at 9.5\,dB squeezing (\( r \approx 1 \)), the 
overlap
between the codewords of the optimal squeezed Fock code (\( n=1 \)) is \( \vert \langle 1_{\text{L}} \vert 0_{\text{L}} \rangle \vert = \cosh(2r)^{-3/2} \approx 0.137 \), whereas our proposed code is exactly orthogonal. This intrinsic orthogonality substantially relaxes the squeezing requirement while ensuring robust and efficient quantum error correction against photon loss and dephasing.

We next evaluate the nonvanishing KL terms for the squeezed cat code,
\begin{equation}
	\vert0_L/1_L\rangle=\frac{1}{\mathcal{N}_{\pm}}\left(\vert\beta,r\rangle\pm\vert-\beta,r\rangle\right),
\end{equation}
and assess its error-correction performance, where  $\vert \beta,r\rangle$ is the squeezed coherent state and the normalization coefficients are $ \mathcal{N}_{\pm}=\sqrt{2\left[1\pm \exp\left(-2e^{2r}\beta^2\right)\right]}$. These codewords are mutually orthogonal. For \(\vert\beta\vert > e^{-r}\) and as \(r \rightarrow \infty\), the relation \(\langle 0_L\vert \hat{A}^{\dagger}_i \hat{A}_j \vert 0_L \rangle = \langle 1_L \vert \hat{A}^{\dagger}_i \hat{A}_j \vert 1_L \rangle\) is approximately satisfied for the error set $\mathbf{E}$.
However, the error basis is not orthogonal, i.e., \(\langle 1_L \vert \hat{a}\hat{n}^m \vert 0_L \rangle\) and \(\langle 1_L \vert \hat{a}^{\dagger}\hat{n}^m \vert 0_L \rangle\) do not vanish. {The nonvanishing terms $\delta_{\hat{A},\hat{A}^{\dagger}} = \langle 1_L | \hat{A} | 0_L \rangle$ and $\langle 1_L | \hat{A}^\dagger | 0_L \rangle$ are listed below.}
\begin{equation}
	\begin{aligned}
	\delta_{\hat{a}^{\dagger},\hat{a}}&=
\frac{\beta\left(1\mp e^{-2e^{2r}\beta^{2}+2r}\right)}{\sqrt{1-e^{-4e^{2r}\beta^{2}}}}, \;\;\;
\delta_{\hat{a}^{\dagger}\hat{n},\hat{n}\hat{a}}=\frac{3\beta e^{-2r}+\beta e^{2r}+4\beta\left(\beta^{2}-1\right)\pm\beta e^{-2e^{2r}\beta^{2}}\left(4e^{6r}\beta^{2}-1+4e^{2r}-3e^{4r}\right)}{4\sqrt{1-e^{-4e^{2r}\beta^{2}}}},\\
\delta_{\hat{a}^{\dagger}\hat{n}^2,\hat{n}^2\hat{a}}&=\frac{1}{16\sqrt{1-e^{-4e^{2r}\beta^{2}}}}\Big[15\beta e^{-4r}+3\beta e^{4r}+8\left(\beta^{2}-1\right)\beta e^{2r}+8\left(5\beta^{2}-3\right)\beta e^{-4r}+2\left(8\beta^{4}-16\beta^{2}+7\right)\beta  \\
&\text{\ensuremath{\pm}}\beta e^{-2e^{2r}\beta^{2}}\left(-16e^{10r}\beta^{4}+40e^{8r}\beta^{2}+8e^{4r}\left(\beta^{2}+3\right)-e^{6r}\left(32\beta^{2}+15\right)-3e^{-2r}+8-14e^{2r}\right)\Big],
	\end{aligned}
	\label{equation_kl}
\end{equation}
{where the \( \pm \) and \( \mp \) subscripts distinguish between \( \hat{A} \) and \( \hat{A}^{\dagger} \), respectively.}
From the above equations, we find that \(\delta_{\hat{a}^{\dagger},\hat{a}} \to \beta\) as \(r \to \infty\) and \(\vert \beta \vert > e^{-r}\), while the other terms cannot simultaneously approach zero. Therefore, the squeezed cat code cannot correct error sets that involve both \(\hat{a}\) and \(\hat{n}^m\) with \(m > 0\). 
 {Since the error sets for single-photon loss and dephasing channels are $[\hat{I}, \hat{a}, \hat{n}]$ and $[\hat{I}, \hat{n}, \hat{n}^2]$, respectively, the squeezed cat code can correct dephasing errors, but exhibits degraded performance under single-photon loss, as the corresponding terms in Eq.~(\ref{equation_kl}) do not vanish.
 	
 	We also include both the numerical and analytical results for $K_{\text{err}}$ of the squeezed cat codes as a function of the parameter $\beta$, for squeezing levels $r = 0.5$, $1$, and $1.5$, under the error set $\mathbf{E}$, as shown in Fig.~\ref{fig:lru_scheme_2d}(c). The excellent agreement between the two confirms the validity of our analytical solutions and further indicates that the squeezed cat code performs poorly under the error set $\mathbf{E}$.}
We present the error correction performance of these three codes in Table \ref{table1}.

\renewcommand{\arraystretch}{1.5}
\begin{table}[h!]
	\centering
	\setlength{\tabcolsep}{3.6mm}{
		\begin{tabular}{c|c|c|c|c}
			\hline
			&Fully Orthogonal & Single-photon loss & Dephasing & Single-photon loss $\&$ Dephasing   \\
			\hline
			Our code & \ding{51} & \ding{51}\ding{51} ($\varpropto e^{-7r}$) & \ding{51}\ding{51} ($\varpropto e^{-7r}$) & \ding{51}\ding{51} ($\varpropto e^{-7r}$) \\
			\hline
			Squeezed cat code & \ding{55} & \ding{55} & \ding{51}& \ding{55}\\
			\hline
			Squeezed Fock code & \ding{55}  &  \ding{51} ($\varpropto e^{-3r}$) & \ding{51} ($\varpropto e^{-3r}$) & \ding{51} ($\varpropto e^{-3r}$)    \\
			\hline
		\end{tabular}
	}
	\caption{Comparison of different squeezed-bosonic codes for squeezing $r\rightarrow\infty$.}
	\label{table1}
\end{table}

\section{Construction of Logical Pauli Operators}
Owing to the rotational relation of the encoded logical states $\vert 0_L\rangle$ and $\vert 1_L\rangle$, a logical bit-flip operation can be implemented via a $\pi/2$ rotation generated by $\hat{X}_L = \exp(-i \frac{\pi}{2} \hat{n})$. In contrast, implementing the logical phase-flip operator is less straightforward and is therefore obtained via numerical optimization. Given the continuous nature of the encoding, we adopt an ansatz inspired by the finite-energy Gottesman–Kitaev–Preskill (GKP) logical Pauli operators. Specifically, the logical Pauli-$Z$ operator is realized as
\begin{equation}
	\hat{Z}_L = \exp[-i \hat{H}_z(\hat{a}, \hat{a}^\dagger)],
	\label{eq_h}
\end{equation}
where $\hat{H}_z$ is an operator parameterized in two alternative forms for numerical construction.

Motivated by the non-Hermitian nature of Pauli operators in the GKP code, we parameterize \( \hat{H}_z \) as
\begin{equation}
	\hat{H}_z(\hat{a}, \hat{a}^\dagger) = \sum_{k,l=0}^{n} \alpha_{kl} \frac{\hat{a}^{\dagger k} \hat{a}^l}{\|\hat{a}^{\dagger k} \hat{a}^l\|_2},
\end{equation}
where \( \alpha_{kl} \) are complex variational parameters and \( \|\cdot\|_2 \) denotes the matrix 2-norm. Realizing an effective non-Hermitian logical Pauli-\( Z \) operator necessitates incorporating an auxiliary system for its construction.

Alternatively, we consider \( \hat{H}_z(\hat{a}, \hat{a}^\dagger) \) in Hermitian form to enable a direct unitary gate acting on the encoded bosonic mode without requiring an auxiliary system.
To this end, we employ a symmetrized ansatz:
\begin{equation}
	\hat{H}_z(\hat{a}, \hat{a}^\dagger) = \sum_{k,l=0}^{n} \left( \alpha_{kl} \frac{\hat{a}^{\dagger k} \hat{a}^l}{\|\hat{a}^{\dagger k} \hat{a}^l\|_2} + \alpha^*_{kl} \frac{\hat{a}^{k} \hat{a}^{\dagger l}}{\|\hat{a}^{\dagger l} \hat{a}^{k}\|_2} \right),
\end{equation}
which guarantees that $\hat{H}_z = \hat{H}_z^\dagger$ by construction. 

In both approaches, the variational parameters $\alpha_{kl}$ are optimized numerically to approximate the desired logical Pauli-$Z$ operator. To this end, we minimize the following loss function:
\begin{equation}
	E = \sum_{u=0,1} \left( \left| \langle u_L \vert \hat{Z}_L \vert u_L \rangle - (-1)^u \right|^2 + \left| \langle u_L \vert \hat{Z}_L^\dagger \vert u_L \rangle - (-1)^u \right|^2 + \left| \langle u_L \vert \hat{Z}_L^\dagger \hat{Z}_L \vert u_L \rangle - 1 \right|^2 \right),
\end{equation}
where the first two terms enforce that $\vert u_L \rangle$ are eigenstates of the operator $\hat{Z}_L$ with eigenvalues $\pm 1$, while the last term ensures that $\hat{Z}_L$ acts unitarily on the logical states. This optimization is performed using the Adam algorithm implemented in the PyTorch Python package.

Numerical simulations reveal that constructing the logical Pauli-\(Z\) operator \(\hat{Z}_L\) using a non-Hermitian generator requires only a small set of operator terms:
$
\{ \hat{I},\ \hat{a}^{\dagger 2},\ \hat{a}^2,\ \hat{a}^\dagger \hat{a},\ \hat{a}^{\dagger 2} \hat{a}^2 \}.
$
This choice balances expressive power and computational efficiency, involving only five complex variational parameters. Despite its compactness, the ansatz achieves a high-fidelity approximation of the logical gate, with a loss function below \(10^{-4}\).
As a representative example at 8\,dB squeezing, the optimized coefficients are found to be
\begin{equation}
\boldsymbol{\alpha}\approx[1.5741 - 0.1206i,\quad 116.2624 - 0.1807i,\quad -53.0023 + 0.1887i,\quad -0.3235 + 19.8160i,\quad 5.9651 - 433.85i].
\end{equation}
In contrast, enforcing Hermiticity to ensure a unitary representation of \( \hat{Z}_L \) requires a larger operator basis. For comparable accuracy (\( E < 10^{-4} \)), we adopt an ansatz with \( n = 6 \) symmetrized terms. The corresponding optimized parameters for the Hermitian case at 8\,dB squeezing are:
\begin{equation}
	\boldsymbol{\alpha} \approx \Re(\boldsymbol{\alpha}) + i\,\Im(\boldsymbol{\alpha}),
\end{equation}
where
\begin{equation}
	\Re(\alpha)=
	\begin{bmatrix}
		0.0 & 6.7115 & 0.29561 & 240.13 & 0.09506 & 9956.3 \\
		6.6332 & 0.0414 & -360.64 & 0.02024 & 868.27 & 0.20038 \\
		-0.2944 & -360.69 & -0.77259 & -11082 & -0.9056 & 118890 \\
		240.35 & 0.02024 & -11082 & 0.19964 & -12945 & 0.0 \\
		-0.01497 & 868.11 & -0.9056 & -12944 & 0.0 & 0.0 \\
		9955.9 & 0.20038 & 11889 & 0.0 & 0.0 & 0.0
	\end{bmatrix},
\end{equation}
\begin{equation}
	\Im(\alpha)=
	\begin{bmatrix}
		0 & 0.22634 & 0.0 & 0.77684 & 0 & 0.43069 \\
		0.22633 & 0.0 & 0.30752 & 0.0 & -0.00143 & 0.0 \\
		0 & 0.31577 & 0 & 0.03513 & 0 & -0.78899 \\
		0.78078 & 0 & 0.03969 & 0.0 & -1.4467 & 0.0 \\
		0.0 & 0.0 & 0.0 & 2.3474 & 0.0 & 0.0 \\
		0.26227 & 0.0 & 0.69542 & 0.0 & 0.0 & 0.0
	\end{bmatrix}.
\end{equation}

\section{Quantum error correction approaches}
We develop error-correction approaches based on the codewords Eq.~(\ref{codewords}). These schemes encompass measurement-based error correction and autonomous error correction techniques. {The error subspaces induced by the loss–dephasing Kraus operators $\hat{A}_1$ and $\hat{A}_2$ are generally \emph{not} mutually orthogonal.  
	To facilitate the construction of recovery operators, we define an equivalent set of Kraus operators $\hat{F}_1$ and $\hat{F}_2$ such that the corresponding error subspaces are orthogonal.  
	This is achieved by diagonalizing the Hermitian matrix
	\begin{equation}
		\boldsymbol{J} = V \boldsymbol{\Lambda} V^\dagger, \quad 
		J_{ij} = \langle u_L | \hat{A}_i^\dagger \hat{A}_j | u_L \rangle,
	\end{equation}
	where $|u_L\rangle$ denotes the logical codeword and $V$ is a unitary operator. Since only the annihilation operator changes the photon-number parity, we have $J_{i3} = J_{3i} = 0$,
	and therefore $
		\hat{F}_3 = \hat{A}_3$ and $\hat{F}_i = \sum^{2}_{k=1} V_{ki} \hat{A}_k$ ($i=1,2$). This ensures that the error subspaces are approximately orthogonal, 
		$
		\langle u_L | \hat{F}_i^\dagger \hat{F}_j | v_L \rangle \approx \delta_{ij}\delta_{uv}
		$
		(a similar construction can be found in Ref.~\cite{10.21468/SciPostPhysLectNotes.70}).} To facilitate the design of recovery unitary operators, we can reformulate Eq.~(\ref{kraus_oper}) as follows:
{		
\begin{equation}
	\hat{\rho}(\tau)\approx\sum_{i=1}^3 \hat{F}_i \hat{\rho} \hat{F}_i^{\dagger}=\sum_{i=1}^3\sum_{kl} V_{ki}V^*_{li}\hat{A}_k \hat{\rho} \hat{A}_l^{\dagger}=\sum_{kl}\delta_{kl}\hat{A}_k \hat{\rho} \hat{A}_l^{\dagger}=\sum_{k}\hat{A}_k \hat{\rho} \hat{A}_k^{\dagger}. 
\end{equation}
}Consequently, we can rewrite the approximate KL condition as{
\begin{equation}
	\hat{P}_L \hat{F}_i^{\dagger} \hat{F}_j \hat{P}_L=	 \sum_{kl}V^*_{ki}V_{lj}\hat{P}_L\hat{A}_k^{\dagger} \hat{A}_l\hat{P}_L\approx\sum_{kl}V^*_{ki}V_{lj}J_{kl}\hat{P}_L=  \Lambda_{ii} \delta_{ij} \hat{P}_L,
\end{equation}
}where $\hat{P}_L=\vert0_L\rangle\langle 0_L\vert+\vert1_L\rangle\langle 1_L\vert$ is the projector operator of the code space.

\begin{figure}[t]
	\centering
	\includegraphics[width=0.98\textwidth]{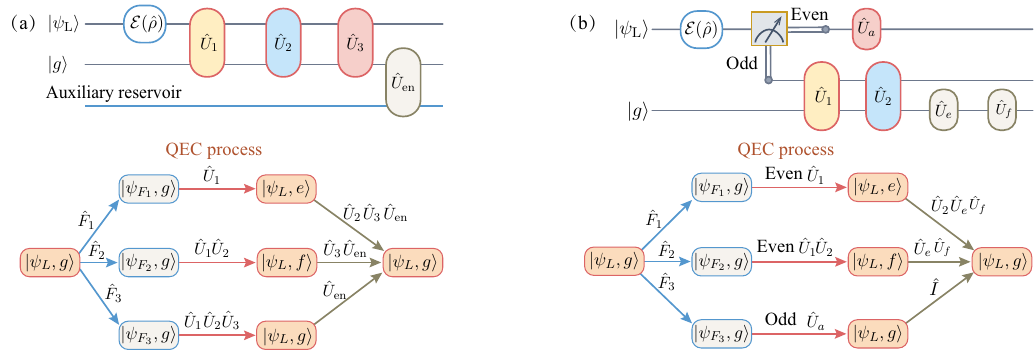}
	\caption{(a)  
		Schematic of autonomous quantum error correction. 		
		The protocol employs a sequence of unitary operations \( \hat{U}_i \), each designed to address errors associated with the operators \( \hat{F}_i \) ($i = 1, 2, 3$). {The errors $\hat{F}_1$ and $\hat{F}_2$ are coherently mapped to the excited states \( \lvert e \rangle \) and \( \lvert f \rangle \) of an ancilla (respectively), while for a parity error $\hat{F}_3$ the ancilla remains in its ground state \( \lvert g \rangle \),
        thereby enabling selective and efficient correction.}	
		An auxiliary reservoir, such as a fast-decaying bosonic mode or a qudit, mediates transitions by transferring the states \( \vert \psi_{\text{L}}, e \rangle \) and \( \vert \psi_{\text{L}}, f \rangle \) to \( \vert \psi_{\text{L}}, g \rangle \) via the unitary \( \hat{U}_{\text{en}} \). Each operation \( \hat{U}_i \) operates exclusively on its designated error subspace.
		(b) Quantum error correction based on parity measurement. If an even parity is detected, indicating the occurrence of an error \(\hat{a}\), the unitary { \(\hat{U}_{a}=\hat{L}^{\dagger}_3+\hat{L}_3+\hat{I}-\hat{P}_{L}-\hat{P}_{F_3}\)} is applied to restore the state. For odd parity detection, unitaries \(\hat{U}_1\) and \(\hat{U}_2\) are employed to recover the encoded states. Additionally, the gates \(\hat{U}_e\) and \(\hat{U}_f\) facilitate transitions from the excited states \(\vert e \rangle\) and \(\vert f \rangle\) to the ground state \(\vert g \rangle\). Here, $\vert\psi_{L}\rangle$ represents the encoded state, while 
		$\vert\psi_{F_i}\rangle$ denotes the error state in the 
		$i$-th error space.}
	\label{fig:error_correction_scheme}
\end{figure}

Here, we design two physical models to implement quantum error correction based on the encoding (\ref{codewords}). The first model involves coupling the encoded mode with a three-level atom, while the second model 
couples the encoded mode to two two-level atoms for error correction.
We design an autonomous quantum error correction protocol and a parity-measurement-based quantum error correction procedure as shown in Fig.~\ref{fig:error_correction_scheme}. We construct three unitary operations $\hat{U}_i$ to recover the errors $\hat{F}_i$, respectively, where the encoding mode couples to an auxiliary system which exhibits at least three non-degenerate energy levels $\vert g\rangle$, $\vert e\rangle$, and $\vert f\rangle$. This ancilla system is coupled to an auxiliary reservoir, which restores the auxiliary qutrit to its ground state $\vert g\rangle$. Moreover, we also engineer a parity-based measurement quantum error correction scheme in Fig.~\ref{fig:error_correction_scheme}~(b). It should be noted that we can realize the quantum gate with different systems. 

\subsection{Ancilla qutrit system}

Here, we consider a bosonic mode coupled with a qutrit auxiliary system. {At the beginning of the cycle, the qutrit is initialized in the ground state $\vert g\rangle$ and the bosonic mode is encoded in $\vert\psi_{\mathrm{L}}\rangle$.
	The error operators $F_i$ only affect the bosonic mode. Therefore, upon the occurrence of an error, the qutrit always remains in the qutrit ground state.
	During the recovery process, the appropriate unitary $\hat{U}_i$ coherently excites the qutrit to $\vert e\rangle$, $\vert f\rangle$, or leaves it in $\vert g\rangle$, providing a one-to-one correspondence between the qutrit state and the error channel for deterministic correction.}  We can construct the following unitary operators to correct the errors,
\begin{equation}
\begin{split}
	\hat{U}_1&=\hat{L}_1\vert e\rangle\langle g\vert+\hat{L}^{\dagger}_1\vert g\rangle\langle e\vert+\left(I-\hat{P}_{L}\right)\vert e\rangle\langle e\vert+\left(I-\hat{P}_{F_{1}}\right)\vert g\rangle\langle g\vert+\vert f\rangle\langle f\vert, \\
	\hat{U}_2&=\hat{L}_2\vert f\rangle\langle g\vert+\hat{L}^{\dagger}_2\vert g\rangle\langle f\vert+\left(I-\hat{P}_{L}\right)\vert f\rangle\langle f\vert+\left(I-\hat{P}_{F_{2}}\right)\vert g\rangle\langle g\vert+\vert e\rangle\langle e\vert,\\
\hat{U}_3&=\left(\hat{L}_3+\hat{L}^{\dagger}_3+\hat{I}-\hat{P}_{L}-\hat{P}_{F_3}\right)\vert g\rangle\langle g\vert+\vert e\rangle\langle e\vert+\vert f\rangle\langle f\vert,	
\end{split}	
\end{equation}
where we have introduced the projection operators on the different error spaces $\hat{P}_{F_i}=\vert 0_{F_i}\rangle\langle 0_{F_i}\vert+\vert 1_{F_i}\rangle\langle 1_{F_i}\vert$ ($i=1, 2, 3$) and the error correction operators  $\hat{L}_i=\vert0_{L}\rangle\langle0_{F_{i}}\vert+\vert1_{L}\rangle\langle1_{F_{i}}\vert$~{($\vert u_{F_i}\rangle=\hat{F}_i\vert u_{F_i}\rangle/\parallel\hat{F}_i\vert u_{F_i}\rangle\parallel$)}.
 Here, the unitaries \(\hat{U}_i\) are applied sequentially, with each of them correcting only the \(i\)-th error without influencing other error subspaces. {Specifically, the first term of $\hat{U}_{1/2}$ corrects the $\hat{F}_{1/2}$ errors while simultaneously exciting the ancilla. In contrast, the remaining terms ensure unitarity and act as the identity when the $\hat{F}_3$ error occurs. 
 	Note that $\hat{U}_3$ is applied after $\hat{U}_1$ and $\hat{U}_2$: its first term corrects the $\hat{F}_3$ error with the ancilla in the ground-state subspace, while the others correspond to the identity in the excited subspace, leaving the recovered information unaffected and ensuring unitarity. } 
 	
 	As a result of these operations, errors are transferred to an auxiliary qutrit: after correcting \(\hat{F}_1\), \(\hat{F}_2\), and \(\hat{F}_3\), the auxiliary qutrit is left in the states \(\vert e\rangle\), \(\vert f\rangle\), and \(\vert g\rangle\), respectively. This allows the identification of the error type through a measurement of the auxiliary qutrit's state, followed by resetting the qutrit to the ground state \(\vert g\rangle\).  

While this method constitutes a semi-autonomous error correction process—owing to the measurement step—a fully autonomous approach can also be implemented. In this case, measurements are avoided, and after error correction, the auxiliary qutrit is coupled to a highly dissipative auxiliary reservoir: a qubit or bosonic mode, as depicted in Fig.~\ref{fig:error_correction_scheme}(a). The quantum gate \(\hat{U}_{\text{en}}\) resets the auxiliary qutrit to the ground state \(\vert g\rangle\), with flexibility in its implementation depending on system requirements. For example, we can design the quantum unitary operation $\hat{U}_{\text{en}}$ as
\begin{equation}
	\hat{U}_{\text{en}}=\exp\left\lbrace-i\varsigma\left[\left(\vert g\rangle\langle e\vert+\vert g\rangle\langle f\vert\right)\hat{b}^{\dagger}+\left(\vert e\rangle\langle g\vert+\vert f\rangle\langle g\vert\right)\hat{b}\right]t\right\rbrace,
\end{equation}
where $\hat{b}$ describes the auxiliary system with a decay rate $\kappa_b$ much larger than the coupling strength and the qutrit decay rate $\kappa_b\gg \varsigma\gg\gamma$. 
Therefore, the entire error correction process can be described as follows
\begin{equation}
	\mathcal{R}\circ\mathcal{E}\left[\hat{\rho}(t)\right]=\hat{U}_{\text{en}}\hat{U}_3\hat{U}_2\hat{U}_1\mathcal{E}[\hat{\rho}(t)]\otimes\vert g\rangle\langle g\vert\hat{U}^{\dagger}_1\hat{U}^{\dagger}_2\hat{U}^{\dagger}_3\hat{U}^{\dagger}_{\text{en}}\approx\hat{\rho}(t)\otimes\vert g\rangle\langle g\vert,
	\label{sequation_10}
\end{equation}
where we have assumed that the initial state of the qutrit is the ground state $\vert g\rangle\langle g\vert$.

{In the parity-measurement-based QEC scheme, the encoded space lies in the odd-parity Fock subspace, and only the error operator \( \hat{a} \) causes a parity change.} Therefore, an alternatively parity measurement can be employed: if an even parity is detected, the corrective operation, 
\begin{equation}
\hat{U}_a=\vert0_{L}\rangle\langle0_{F_3}\vert+\vert1_{L}\rangle\langle1_{F_3}\vert+\vert0_{F_3}\rangle\langle0_{L}\vert+\vert1_{F_3}\rangle\langle1_{L}\vert+\hat{I}-\hat{P}_{L}-\hat{P}_{F_3}, \label{ua}
\end{equation}
	is applied; otherwise, the sequence \(\hat{U}_1\hat{U}_2\hat{U}_e\hat{U}_f\) is implemented. This process results in the final quantum state \(\hat{\rho} \otimes \vert g\rangle\langle g\vert\), as shown in Fig.~\ref{fig:error_correction_scheme}(b). Here, \(\hat{U}_e\) and \(\hat{U}_f\) restore the auxiliary system to its ground state, with their explicit forms defined as follows:
\begin{equation}
	\begin{split}
\hat{U}_e=\vert e\rangle\langle g\vert+\vert g\rangle\langle e\vert+\vert f\rangle\langle f\vert,~~~~~
\hat{U}_f=\vert f\rangle\langle g\vert+\vert g\rangle\langle f\vert+\vert e\rangle\langle e\vert.
\end{split}
\end{equation}

Therefore, we can use the following equation to describe the quantum error correction process
\begin{equation}
	\mathcal{R}_{m}\circ\mathcal{E}\left[\hat{\rho}(t)\right]=\hat{U}_a\hat{F}_3\hat{\rho}(t)\hat{F}^{\dagger}_3\hat{U}^{\dagger}_a+\sum^2_{i=1}\hat{U}_f\hat{U}_e\hat{U}_2\hat{U}_1\hat{F}_i\hat{\rho}(t)\otimes\vert g\rangle\langle g\vert\hat{F}^{\dagger}_i\hat{U}^{\dagger}_1\hat{U}^{\dagger}_2\hat{U}^{\dagger}_e\hat{U}^{\dagger}_f\approx \hat{\rho}\otimes\vert g\rangle\langle g\vert.
\end{equation}


\begin{figure}[htp]
	\centering
	\includegraphics[width=1.02\textwidth]{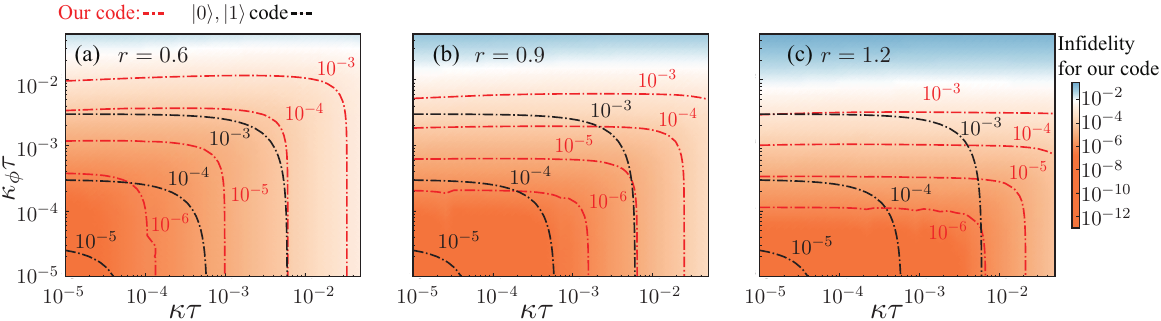}
	\caption{Near-optimal channel infidelity obtained using the Petz recovery channel, shown as a function of $\kappa\tau$ and $\kappa_{\phi}\tau$ for three squeezing parameters: (a) $r = 0.6$, (b) $r = 0.9$, and (c) $r = 1.2$. In each panel, the corresponding $\vert 0\rangle$, $\vert 1\rangle$ encoding channel infidelity contour lines are superimposed to highlight the parameter regions where the Petz-corrected channel surpasses the $\vert 0\rangle,~\vert 1\rangle$ code.
	}
	\label{fig:lru_scheme_2}
\end{figure}
\subsection{Two ancilla qubit systems}
\label{Two_ancilla_qubit_systems}
The auxiliary system can also be designed by two two-level systems. {To extend the method to a two-qubit system, it suffices to replace the qutrit states for the joint two-qubit states: \( \lvert g \rangle \rightarrow \lvert g_1, g_2 \rangle \), \( \lvert e \rangle \rightarrow \lvert e_1, g_2 \rangle \), and \( \lvert f \rangle \rightarrow \lvert g_1, e_2 \rangle \), from which the QEC protocol follows directly.}
Here, we analyze the implementation of autonomous error correction and measurement-based error correction using two auxiliary two-level systems. We only need to replace the quantum gates \( \hat{U}_i \) with the following gates to achieve the desired functionality
\begin{equation}
	\begin{split}
\hat{U}_1=&\left[\hat{L}_1\vert e_{1}\rangle\langle g_{1}\vert+\left(\hat{I}-\hat{P}_{F_{1}}\right)\vert g_{1}\rangle\langle g_{1}\vert+\hat{L}^{\dagger}_1\vert g_{1}\rangle\langle e_{1}\vert+\left(I-\hat{P}_{L}\right)\vert e_{1}\rangle\langle e_{1}\vert\right]\vert g_{2}\rangle\langle g_{2}\vert+\vert e_{2}\rangle\langle e_{2}\vert, \\		
\hat{U}_2=&\left[\hat{L}_2\vert e_{2}\rangle\langle g_{2}\vert+\left(\hat{I}-\hat{P}_{F_{2}}\right)\vert g_{2}\rangle\langle g_{2}\vert+\hat{L}^{\dagger}_2\vert g_{2}\rangle\langle e_{2}\vert+\left(I-\hat{P}_{L}\right)\vert e_{2}\rangle\langle e_{2}\vert\right]\vert g_{1}\rangle\langle g_{1}\vert+\vert e_{1}\rangle\langle e_{1}\vert,\\
\hat{U}_3=&\left(\hat{L}_3+\hat{L}^{\dagger}_3+\hat{I}-\hat{P}_L-\hat{P}_{F_3}\right)\vert e_1g_2\rangle\langle g_1g_2\vert+\vert e_{1}e_{2}\rangle\langle e_{1}e_{2}\vert+\vert g_{1}e_{2}\rangle\langle g_{1}e_{2}\vert+\vert g_{1}g_{2}\rangle\langle e_{1}g_{2}\vert,
\end{split}
\label{two_atom_gate}
\end{equation}

For the measurement-based quantum error correction, we only need to replace the quantum gates $\hat{U}_e$ and $\hat{U}_f$ by the Pauli operators $\hat{\sigma}_x$ of the two  ancilla qubits and the design of the quantum gate $\hat{U}_{\text{en}}$ is similar with Eq.~(\ref{sequation_10}). Therefore, we can achieve the quantum error correction with the two  ancilla qubit model. 
 By performing a parity measurement, we enable QEC. Specifically, if the even parity is measured, we apply the recovery operation \(\hat{U}_a\) in Eq.~(\ref{ua}) to correct errors caused by $\hat{a}$. For odd parity outcomes, the errors are corrected using the operation $\hat{U}_2\hat{U}_1$. After the encoded information is successfully corrected, the ancilla qutrit states are transferred into distinct configurations. Finally, a quantum gate $\hat{U}_r$ is employed to reset the two ancilla qubits to their ground state $\vert g\rangle$.
 The entire parity-based QEC process can be formally represented as follows
 \begin{equation}
 	\begin{split}
 		\mathcal{R}_{m}\circ\mathcal{E}\left[\hat{\rho}_{L}\right]\approx&\sum^2_{i=1}\hat{U}_{\text{r}}\hat{U}_2\hat{U}_1\hat{F}_i\hat{\rho}_L\otimes\vert g\rangle\langle g\vert\hat{F}^{\dagger}_i\hat{U}^{\dagger}_1\hat{U}^{\dagger}_2\hat{U}^{\dagger}_{\text{r}} \\
 		&	+\hat{U}_a\hat{F}_3\hat{\rho}_L\otimes\vert g\rangle\langle g\vert\hat{F}^{\dagger}_3\hat{U}^{\dagger}_a
 		\\
 		&
 		\approx \hat{\rho}_L\otimes\vert g\rangle\langle g\vert.
 	\end{split}
 \end{equation}

{
\subsection{Near-optimal recovery channel}
Here, we consider a near-optimal recovery channel, the Petz recovery map $\mathcal{R}_p[\cdot]$, which serves as a theoretically motivated benchmark for QEC~\cite{Kwon2022Jan,Li2025May}.
The Petz recovery map is defined as
\begin{equation}
	\mathcal{R}_p\!\left[\hat{\rho}(\tau)\right]
	= \hat{P}_L\, \mathcal{E}^{\dagger}\!\left[
	\mathcal{E}^{-1/2}\!\left(\hat{P}_L\right)\,
	\hat{\rho}(\tau)\,
	\mathcal{E}^{-1/2}\!\left(\hat{P}_L\right)
	\right]\hat{P}_L ,
\end{equation}
where $\mathcal{E}^{\dagger}$ denotes the adjoint (dual) superoperator of $\mathcal{E}$ with respect to the Hilbert--Schmidt inner product. Here, $\hat{\rho}(\tau)=\mathcal{E}(\hat{\rho}_0)$, where $\hat{\rho}_0$ denotes the initial encoded logical state at time $t_0$. We evaluate the channel infidelity of the Petz recovery channel within the error-correction subspace using a superoperator formalism based on the master equation [Eq.~(\ref{master_equation})], rather than an approximate Kraus-operator description of the noise. The calculations employ sparse-matrix techniques and are accelerated on GPUs~\cite{Mercurio2025Sep}.
The corresponding results are shown in Fig.~\ref{fig:lru_scheme_2}. As is evident, this near-optimal recovery channel already enables efficient QEC at the modest squeezing amplitude $r=0.6$.
With increasing squeezing, the correction performance against single-photon loss is further enhanced, whereas the correction of dephasing errors is slightly reduced.
This trade-off does not degrade the overall error-correction performance, since in most relevant experimental platforms the photon-loss rate dominates over the dephasing rate.
This behavior arises from competing effects.
Increasing the squeezing improves the satisfaction of the KL conditions, thereby enhancing protection against photon-loss errors.
At the same time, stronger squeezing increases the average photon number, which raises the likelihood of multiple error events.
Similarly, as $\kappa_{\phi}\tau$ increases, the validity of the truncated error set $\mathbf{E}$ used in the recovery construction gradually deteriorates.
Despite these limitations, our encoding exhibits near-vanishing infidelity for the combined loss–dephasing channel around $\kappa\tau \approx \kappa_{\phi}\tau\leq 10^{-3}$.
In this regime, the corresponding gain relative to the $\vert0\rangle,~\vert1\rangle$ is strongly enhanced.
}


\section{Experimental proposal for quantum error correction}
We now propose a physical implementation of our quantum error correction approach, where quantum gates are designed using optimal quantum control via the gradient ascent pulse engineering (GRAPE) method. Specifically, we consider a superconducting 3D cavity, hosting the bosonic encoding, dispersively  coupled to an ancilla transmon qutrit enabling control of the bosonic mode. The Hamiltonian of the cavity-transmon system is expressed as  
\begin{equation}
\begin{aligned}
	\hat{H}  & =\omega_s\hat{a}^{\dagger}\hat{a}+\omega_{ge}|e\rangle\langle e|+\omega_{gf}|f\rangle\langle f| -\chi_{\mathrm{e}}|e\rangle\langle e| \hat{a}^{\dagger} \hat{a}-\chi_{f}|f\rangle\langle f|\hat{a}^{\dagger}\hat{a}  +\Omega_{d}(t)\hat{a}^{\dagger}e^{-i\omega_dt}+\Omega^*_{d}(t)\hat{a}e^{i\omega_dt},
\end{aligned}
\end{equation}
where \(\vert g\rangle, \vert e\rangle, \vert f\rangle\) denote the states of the ancilla transmon qutrit, \(\hat{a}\) is the photon annihilation operator of the cavity with the frequency $\omega_s$, and \(\chi_{e/f}\) represent the dispersive interaction strengths. \(\Omega_{d}(t)\) is the classical complex driving amplitude of the cavity driving field with frequency  \(\omega_d\). The transition frequencies for the transmon between the $g\leftrightarrow e$ and $g\leftrightarrow f$ states are \(\omega_{ge}\) and \(\omega_{gf}\), respectively.
This coupled cavity-transmon model has been extensively investigated and reliably demonstrated in recent experiments, supporting significant progress in quantum error correction~\cite{Lescanne2020May}, quantum gate implementation~\cite{Kumar2023Mar}, photon blockade~\cite{Wang2022Jun}, and quantum control~\cite{Eickbusch2022Dec}.

In the rotating frame of the drive and transmon transition frequencies, the time-dependent Hamiltonian becomes  
\begin{equation}
	\hat{H}_{I} =-\chi_{\mathrm{e}}|e\rangle\langle e| \hat{a}^{\dagger} \hat{a}-\chi_{f}|f\rangle\langle f|\hat{a}^{\dagger}\hat{a}  +\Omega_{q}(t)\hat{q}+\Omega_{p}(t)\hat{p},
\end{equation}
where \(\omega_d \approx \omega_s\) is assumed. The complex classical driving amplitude is expressed as \(\sqrt{2}\Omega_d(t) = \Omega_q(t) + i\Omega_p(t)\), with \(\Omega_q(t)\) and \(\Omega_p(t)\) representing the effective real drives amplitudes along the position and momentum quadratures, respectively.
Through a sequence of carefully designed control pulses, all operations on the logical qubit are implemented based on the dispersive interaction between the ancilla and the oscillator.

Note that in our approach, control is applied only to the encoded mode. In principle, however, a variety of control Hamiltonians can be considered, including direct control of the qutrit and its interaction with the bosonic mode, which may further improve the precision. Since the method for optimizing the control fields remains unchanged, we present the encoded-mode control as a representative example. In theory, arbitrary quantum gates for this system can be realized via such optimal control. The dispersive interaction strengths \(\chi_{e}\) and \(\chi_{f}\) can be freely chosen within the range of current experimental capabilities. 
	
	For simplicity, our numerical simulations are performed in units of $\chi$, with $\chi_{e} \approx \chi_{f} = 1$; however, this assumption is not essential for implementing the control scheme. During the GRAPE-based simulation, we neglect noise in both the bosonic mode and the qutrit, as the control Hamiltonian \(\hat{H}_c\) describes a purely coherent control and not designed to suppress noise. In practical gate implementation, the influence of noise can be accounted for by incorporating it into the system dynamics.



\begin{figure}[ht]
	\centering
	\includegraphics[width=0.52\textwidth]{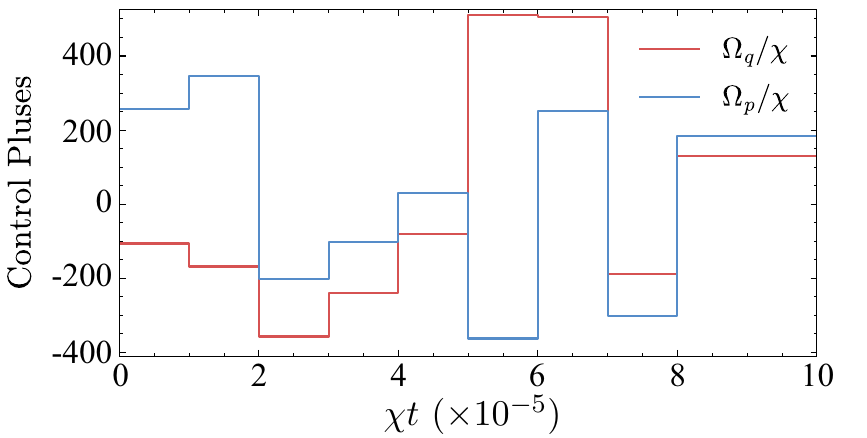}
	\caption{(Color online) Control pulses obtained from the GRAPE algorithm. The control field is discretized into 10 segments with a total evolution time of \( T\chi = 10^{-4} \). Other choices of control Hamiltonians are also feasible and can achieve comparable gate performance.}
	\label{fig:control}
\end{figure}

We employ the QuTiP Python library~\cite{Johansson2012Aug,Johansson2013Apr,BibEntry2024Dec} to numerically implement the GRAPE algorithm and present the optimized control pulses in Fig.~\ref{fig:control}. The resulting gate fidelity between the target unitary \( \hat{U}=\hat{U}_3\hat{U}_2\hat{U}_1 \) and the simulated evolution \( \hat{U}_T \) at time \( T \) exceeds 0.99. {Here, we use the gate fidelity defined as $F=\vert\text{Tr}[U^{\dagger}V]\vert/{d}$, where $U=\hat{U}_3\hat{U}_2\hat{U}_1$ and $V$ denote the target and implemented unitary operations, respectively, and $d$ is the Hilbert-space dimension.}{
 A second transmon qubit can be incorporated in the model in order to realize the protocol described at the end of the previous section. Note that in this case both transmon qubits need to be truncated to their lower two levels.} As the procedure closely mirrors that of the two ancilla qubit-based implementation, we omit further details here.  

\bibliography{supplemental}